\documentclass[%
 reprint,
 superscriptaddress,
 nofootinbib,
 amsmath,amssymb,
 aps,
 prl,
 floatfix
]{revtex4-2}

\usepackage{graphicx}
\usepackage{dcolumn}
\usepackage{bm}
\usepackage{xcolor}
\usepackage[pdftex,colorlinks,linkcolor = blue,urlcolor  = blue,citecolor = blue]{hyperref}

\begin{document}

\preprint{APS/123-QED}

\title{Comprehensive Test of the Brink-Axel Hypothesis in the Energy Region of the Pygmy Dipole Resonance}

\author{M.~Markova}
\email{maria.markova@fys.uio.no}
\affiliation{Department of Physics, University of Oslo, N-0316 Oslo, Norway}

\author{P.~von Neumann-Cosel}%
 \email{vnc@ikp.tu-darmstadt.de}
\affiliation{%
 Institut f\"{u}r Kernphysik, Technische Universit\"{a}t Darmstadt, D-64289 Darmstadt, Germany
}%

\author{A.~C.~Larsen}
\email{a.c.larsen@fys.uio.no}
\affiliation{Department of Physics, University of Oslo, N-0316 Oslo, Norway}

\author{S.~Bassauer}%
\affiliation{%
 Institut f\"{u}r Kernphysik, Technische Universit\"{a}t Darmstadt, D-64289 Darmstadt, Germany
}%

\author{A.~G\"{o}rgen}
\affiliation{Department of Physics, University of Oslo, N-0316 Oslo, Norway}

\author{M.~Guttormsen}
\affiliation{Department of Physics, University of Oslo, N-0316 Oslo, Norway}

\author{F.~L.~Bello Garrote}
\affiliation{Department of Physics, University of Oslo, N-0316 Oslo, Norway}

\author{H. C.~Berg}
\affiliation{Department of Physics, University of Oslo, N-0316 Oslo, Norway}

\author{M.~M.~Bj{\o}r{\o}en}
\affiliation{Department of Physics, University of Oslo, N-0316 Oslo, Norway}

\author{T.~Dahl-Jacobsen}
\affiliation{Department of Physics, University of Oslo, N-0316 Oslo, Norway}

\author{T.~K.~Eriksen}
\affiliation{Department of Physics, University of Oslo, N-0316 Oslo, Norway}

\author{D.~Gjestvang}
\affiliation{Department of Physics, University of Oslo, N-0316 Oslo, Norway}

\author{J.~Isaak}%
\affiliation{%
 Institut f\"{u}r Kernphysik, Technische Universit\"{a}t Darmstadt, D-64289 Darmstadt, Germany
}%

\author{M.~Mbabane}
\affiliation{Department of Physics, University of Oslo, N-0316 Oslo, Norway}

\author{W.~Paulsen}
\affiliation{Department of Physics, University of Oslo, N-0316 Oslo, Norway}

\author{L.~G.~Pedersen}
\affiliation{Department of Physics, University of Oslo, N-0316 Oslo, Norway}

\author{N.~I.~J.~Pettersen}
\affiliation{Department of Physics, University of Oslo, N-0316 Oslo, Norway}

\author{A.~Richter}%
\affiliation{%
 Institut f\"{u}r Kernphysik, Technische Universit\"{a}t Darmstadt, D-64289 Darmstadt, Germany
}%

\author{E.~Sahin}
\affiliation{Department of Physics, University of Oslo, N-0316 Oslo, Norway}

\author{P.~Scholz}
\affiliation{Institut f\"{u}r Kernphysik, Universit\"{a}t zu K\"{o}ln, D-50937 K\"{o}ln, Germany}
\affiliation{Department of Physics, University of Notre Dame, Indiana 46556-5670, USA}

\author{S.~Siem}
\affiliation{Department of Physics, University of Oslo, N-0316 Oslo, Norway}

\author{G.~M.~Tveten}
\affiliation{Department of Physics, University of Oslo, N-0316 Oslo, Norway}

\author{V.~M.~Valsdottir}
\affiliation{Department of Physics, University of Oslo, N-0316 Oslo, Norway}

\author{M.~Wiedeking}
\affiliation{Department of Subatomic Physics, iThemba LABS, Somerset West 7129, South Africa}
\affiliation{School of Physics, University of the Witwatersrand, Johannesburg 2050, South Africa}

\author{F.~Zeiser}
\affiliation{Department of Physics, University of Oslo, N-0316 Oslo, Norway}

\date{\today}

\begin{abstract}
The validity of the Brink-Axel hypothesis, which is especially important for numerous astrophysical calculations, is addressed for $^{116,120,124}$Sn below the neutron separation energy by means of three independent experimental methods. The $\gamma$-ray strength functions (GSFs) extracted from primary $\gamma$-decay spectra following charged-particle reactions with the Oslo method and with the Shape method demonstrate excellent agreement with those deduced from forward-angle inelastic proton scattering at relativistic beam energies. 
In addition, the GSFs are shown to be independent of excitation energies and spins of the initial and final states.
The results provide the critical test of the generalized Brink-Axel hypothesis in heavy nuclei, demonstrating its applicability in the energy region of the pygmy dipole resonance.  
\end{abstract}

\maketitle

\textit{Introduction.}$-$
Gamma-ray  strength functions (GSFs) describe the average $\gamma$ decay and absorption probability of nuclei as a function of $\gamma$ energy.
Besides their genuine interest and importance for basic nuclear physics, they are required for applications in astrophysics \cite{Arnould2020}, reactor design \cite{Chadwick2011}, and waste transmutation \cite{Salvatore2011} based on the application of the statistical nuclear reaction theory.
A particular example is large-scale reaction network calculations of neutron capture reactions in the 
$r$-process nucleosynthesis. 
Accordingly, there are considerable efforts to collect data on the GSF in many nuclei \cite{Goriely2019a} and extract systematic parameterization \cite{Goriely2019b} which allows extrapolation to unknown, exotic cases.

Although all electromagnetic multipoles can in principle contribute, the GSF is dominated by  $E1$ radiation with smaller contributions from $M1$ strength.
Above particle threshold it is dominated by the isovector giant dipole resonance (IVGDR), but at lower excitation energies the situation is complex.
In nuclei with neutron excess one observes the formation of the pygmy dipole resonance (PDR) \cite{Savran2013,Bracco2019} located on the low-energy tail of the IVGDR.
Although the detailed structure of the PDR is under debate, it is commonly believed that its strength is related to the magnitude of neutron excess. 
As the $r$-process involves nuclei with extreme neutron-to-proton ratios, the impact of low-energy $E1$ strength on the $(n,\gamma)$ reaction rates and the resulting $r$-process abundances can be significant \cite{Goriely1998,Goriely2004,Litvinova2009,Daoutidis2012}.

The GSFs used in large-scale astrophysical network calculations of the 
$r$-process \cite{Wiescher2012} are based on model calculations of ground state photoabsorption.   
Their application requires the validity of the Brink-Axel (BA) hypothesis \cite{Brink1955,Axel1962}, which in its generalized form states that the GSF is independent of the energies, spins, and parities of the initial and final states and depends on the $\gamma$ energy only.
However, recent theoretical studies \cite{Misch2014,Johnson2015,Hung2017} put that into question, demonstrating that strength functions of collective modes built on excited states generally do show dependence on the excitation energy.
Shell-model calculations in light nuclei \cite{Johnson2015}  found $E1$ strength functions approximately independent of excitation energy consistent with the BA hypothesis, but it remains an open question whether this can be generalized for heavier nuclei. 

Because of the importance for astrophysical applications, there are many recent experimental studies in the low-energy regime with controversial results, claiming either confirmation \cite{Guttormsen2016,Martin2017,Campo2018,Scholz2020} or violation \cite{Angell2012,Isaak2013,Netterdon2015,Isaak2019} of the BA hypothesis.
Possible non-statistical $\gamma$-width distributions observed in $s$- and $p$-wave neutron capture experiments \cite{Koehler2013} would also represent proof against the BA hypothesis \cite{Fanto2020}. 

There are two major sources of GSF data \cite{Goriely2019a}.
One class of experiments determines the ground-state photoabsorption by measuring the subsequent $\gamma$ \cite{Pietralla2019} or neutron \cite{Berman1975} decay. 
Alternatively, the primary $\gamma$ decay distribution is extracted in light-ion induced compound reactions (the so-called Oslo method \cite{Guttormsen1996, Guttormsen1987, Schiller2000, Larsen11}).
Experiments measuring particle or $\gamma$ decay are limited to the excitation energy region above or below the neutron threshold, respectively.   
In principle, a comparison of $(\gamma,\gamma^\prime)$ and Oslo experiments for the same nucleus should provide a test of the validity of the BA hypothesis~\cite{Renstrom_PhysRevC.93.064302}, but is complicated by the assumptions necessary to extract the GSF. 
For $(\gamma,\gamma')$ experiments with broad bremsstrahlung beams, one needs to model the experimentally inaccessible ground state branching ratios and the  significant contributions to the spectra due to atomic scattering.
The analysis of Oslo-type data is based on the validity of the BA hypothesis, and assumptions have to be made about 
the intrinsic spin distribution and the reaction-dependent spin population.

In such comparisons, possible violations of the BA hypothesis have been observed.  
In heavy deformed nuclei at excitation energies of $2$--$3$~MeV the GSF is dominated by the orbital $M1$ scissors mode \cite{Heyde2010}.
Larger strengths have been reported in most $\gamma$ decay experiments (see, e.g., Refs.~\cite{Krticka2004,Guttormsen2012} than found in the $(\gamma,\gamma^\prime)$ experiments \cite{Enders2005}, but the results strongly depend on the assumed form of the $E1$ strength function in this energy range \cite{Kroll2013}. 
At even lower energies ($< 2$ MeV), a general increase of the GSF (called upbend) is seen in Oslo-type experiments \cite{Voinov2004,Larsen2017}. 
No corresponding strength can be present in ground-state absorption experiments on even-even nuclei due to the pairing properties of the nuclear force, which lead to the absence of levels at low excitation energies.
In the energy region near neutron threshold, a non-statistical decay behavior of the PDR has been reported \cite{Romig2015,Loeher2016}.
The possible impact of the PDR on the GSF extraction is expected to be largest in magic and semi-magic nuclei because of the reduced level density and a shift of part of the strength towards lower energies due to $K$ splitting of the IVGDR in deformed nuclei.

Here we present a benchmark study allowing a test of the BA hypothesis in the energy region of the PDR taking advantage of recent experimental progress, which overcomes most of the problems discussed above.
First, a method for the measurement of $E1$ strength distributions -- and thereby the $E1$ part of the GSFs -- in nuclei from about 5 to 25 MeV has been developed using relativistic Coulomb excitation in inelastic proton scattering at energies of a few hundred MeV and scattering angles close to $0^\circ$ \cite{vonNeumannCosel2019}.
Such data also permit extraction of the $M1$ part of the GSF due to spin-flip excitations \cite{Birkhan2016}, which energetically overlaps with the PDR strength.
Second, a new system for the measurement of $\gamma$ emission in Oslo experiments based on large-volume LaBr$_3$(Ce) detectors
allows qualitatively new tests of the BA hypothesis as described below including resolved coincidence studies of decay to the ground state and low-lying excited states. 

Combining data from the two methods allows for testing the generalized BA hypothesis with respect to the energy and spin independence of initial and final states in the PDR region.
Here we present a case study for $^{116,120,124}$Sn. 
The choice is based on the following considerations.
(i) Data for $E1$ and $M1$ strength distributions in the stable Sn isotopes from $(p,p^\prime)$ experiments at 295 MeV have recently become available \cite{Bassauer2020a,Bassauer2020b} and found to agree well with the GSF above threshold deduced from latest $(\gamma,n)$ experiments by Utsunomiya {\it et al.}~\cite{Utsunomiya2011}.
(ii) The isotopes have high neutron threshold energies providing a large overlap region between the GSFs deduced from the $(p,p^\prime)$ and the Oslo experiments. 
(iii) While their low-energy structure is very similar, the GSFs of the Sn isotopes show a distinct dependence on neutron excess 
in the PDR region~\cite{Bassauer2019}. 

\textit{Experimental details and data analysis.}$-$The inelastic proton scattering experiments and the methods to extract the $E1$ and $M1$ contributions to the GSF are described in detail in Ref.~\cite{Bassauer2020b}. 
The $^{116}$Sn experiment at the Oslo Cyclotron Laboratory (OCL) has  previously been reported in Refs.~\cite{Toft2010,toft2011}. 
A 38 MeV $^3$He beam was used to produce $^{116}$Sn nuclei via the $^{117}$Sn($^{3}$He,$\alpha\gamma$) reaction, where the charged particles were measured with eight collimated Si detectors at 45$^\circ$ and the $\gamma$ rays with the NaI(Tl) array CACTUS~\cite{guttormsen_CACTUS_1990}.
\begin{figure}
\includegraphics[width=1.0\columnwidth]{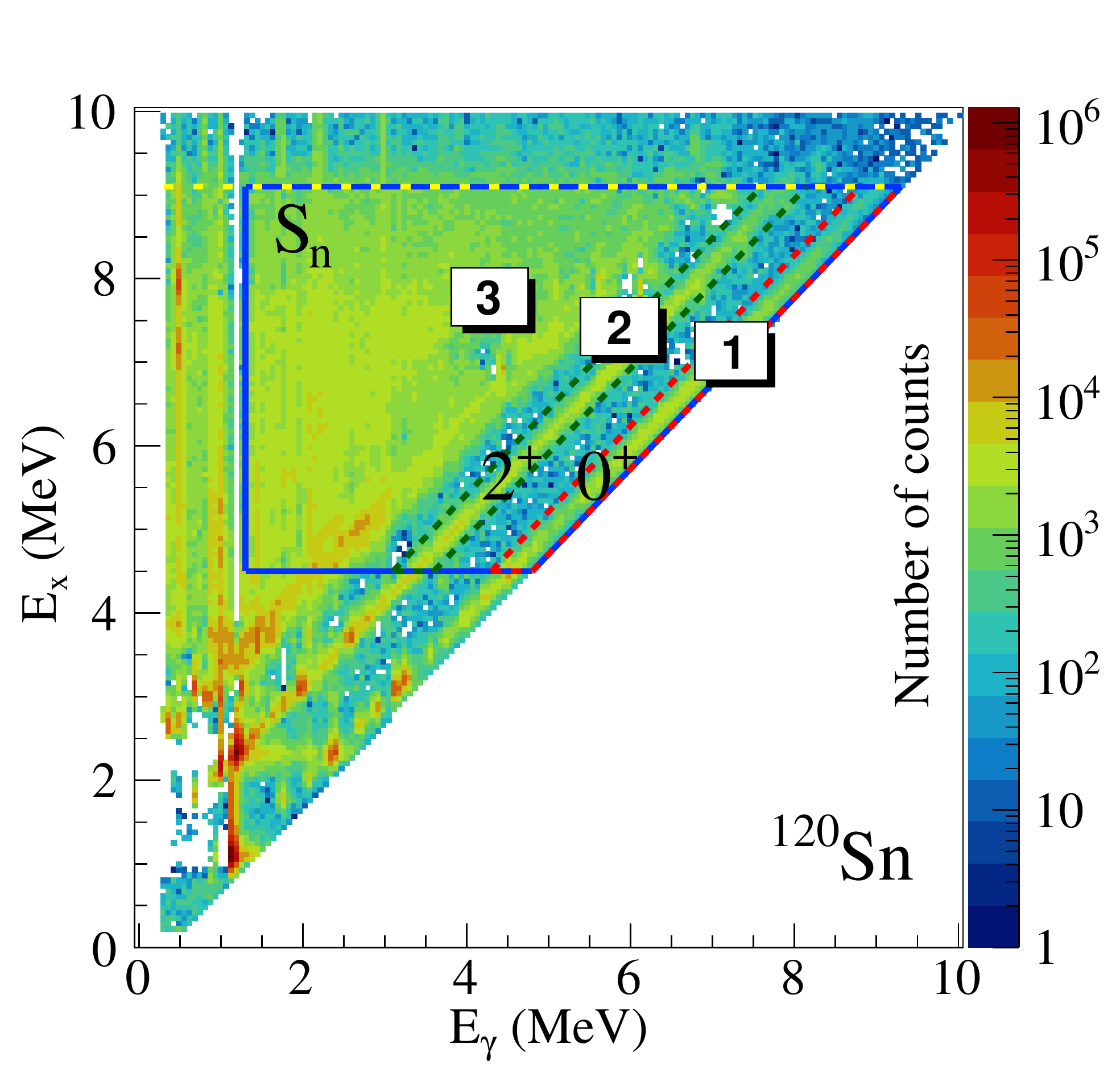}
\caption{\label{fig:fgmatrix} 
Experimental primary $\gamma$-ray matrix $P(E_{\gamma},E_{\rm x})$, Eq.~(\ref{eq:ansatz}), for $^{120}$Sn. 
The yellow dashed line indicates the neutron threshold $S_n$, while the dashed red and blue lines (regions 1 and 2) 
confine transitions to the ground state and the first excited $J^{\pi}=2^+$ state at $E_{\rm x}=1.171$ MeV. 
The solid blue lines (region 3) mark the region 4.5 MeV $\leq E_{\rm x} \leq 9.1$ MeV, $E_{\gamma}\geq 1.3$ MeV used for the Oslo method analysis. }
\end{figure}
%
\begin{figure*}
\includegraphics[width=1.0\textwidth]{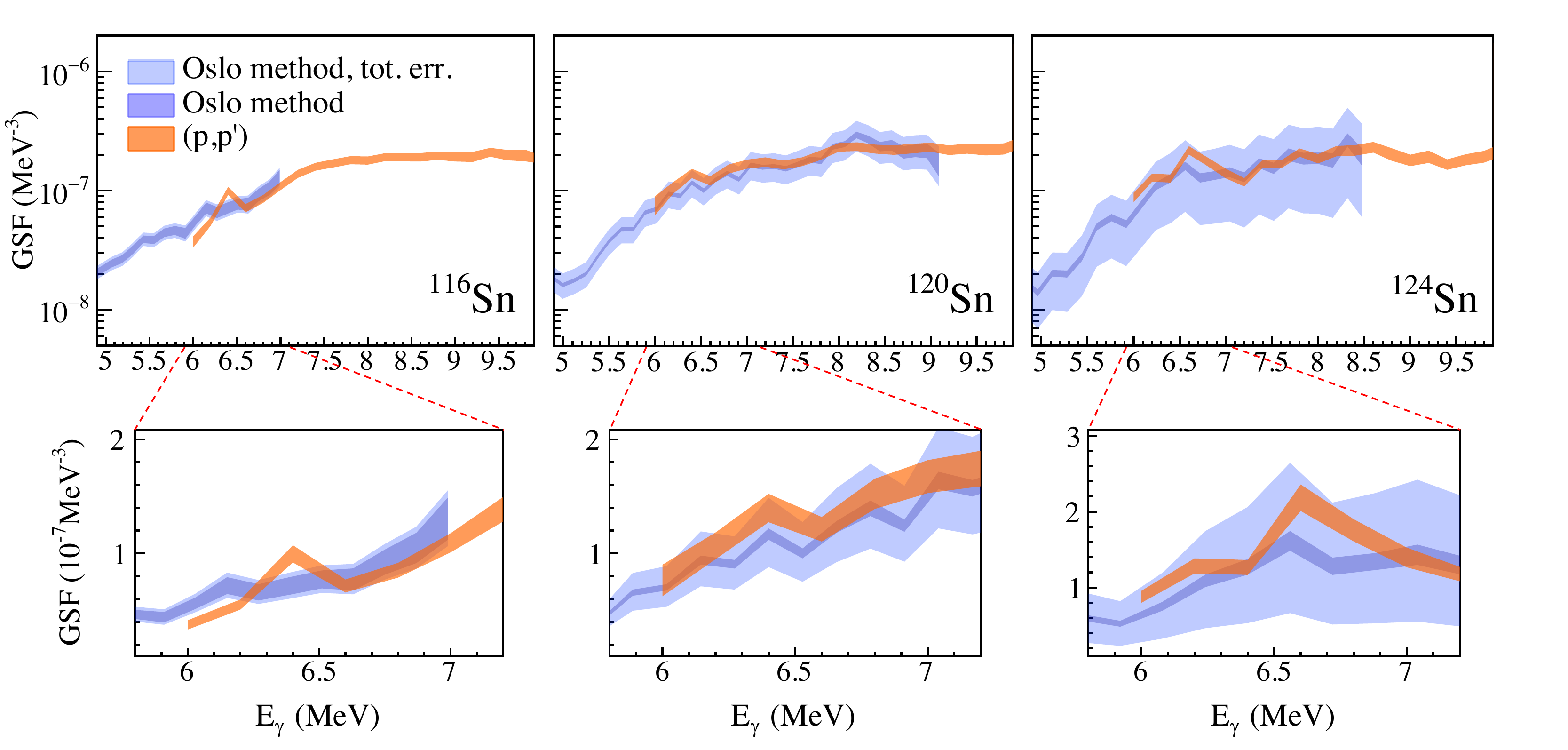}
\caption{\label{fig:gammaSF} Comparison of the GSFs for $^{116,120,124}$Sn obtained from the Oslo method (blue) and from the ($p,p^\prime$) experiments~\cite{Bassauer2020b} (orange). 
The total error bands for the Oslo method (light blue) are asymmetric and include all uncertainties. 
The dark blue band represents   statistical and systematic uncertainties from the unfolding and the extraction of primary $\gamma$ rays. 
The $E_\gamma$ bin widths are 128 keV and 200 keV for the Oslo data and the ($p,p^\prime$) measurements respectively.
}
\end{figure*}

We provide here a brief description of the $^{120,124}$Sn experiments at OCL. 
A 16-MeV proton beam of intensity $I = 3- 4$ nA provided by the MC-35 Scanditronix cyclotron impinged on self-supporting targets of $^{120,124}$Sn. 
The target thicknesses and enrichments were 2.0 mg/cm$^2$, 99.6\% ($^{120}$Sn) and 0.47 mg/cm$^2$, 95.3\% ($^{124}$Sn).
The reactions of interest were $^{120,124}$Sn($p,p^\prime\gamma$).
The targets were placed in the center of the Oslo SCintillator ARray (OSCAR)~\cite{Ingeberg2020,Zeiser2020}, consisting of 30 cylindrical LaBr$_3$(Ce) $\gamma$-ray detectors of size $3.5" \times 8.5"$ mounted on a truncated icosahedron frame. 
Charged particles were registered with 64 Si particle $\Delta E-E$ telescopes (SiRi)~\cite{Guttormsen2011}, covering angles  $126^\circ-140^\circ$. 
The energy resolution of 
OSCAR is $\approx 2.7\%$ at $E_\gamma = 662$~keV.
The front-end of the LaBr$_3$(Ce) crystals were placed 16 cm from the center of the target. 
Particle-$\gamma$ coincidences were recorded using XIA digital electronics \cite{xia}.  
Approximately $5.3 \times10^7$ and $1.3\times10^7$ proton-$\gamma$ coincidences were measured in the excitation energy range up to the neutron thresholds for $^{120}$Sn and $^{124}$Sn, respectively.

The proton energy deposited in the SiRi telescopes was transformed to initial excitation energy $E_{\rm x}$ in the residual nucleus using the reaction kinematics, and the data were arranged in an $E_{\rm x}$ vs.\ $\gamma$-ray energy matrix. 
The $\gamma$-ray spectra for each $E_x$ bin were unfolded with the technique described in Ref.~\cite{Guttormsen1996} using the response function of the OSCAR detectors~\cite{response_function}.
The distribution of primary $\gamma$ rays (the first emitted $\gamma$ rays in the decay cascades) for each $E_{\rm x}$ bin was obtained through an iterative subtraction method~\cite{Guttormsen1987}. 
The resulting primary $\gamma$-ray matrix for the example of $^{120}$Sn is displayed in Fig.~\ref{fig:fgmatrix}.

With the primary $\gamma$-ray matrix $P(E_\gamma,E_{\rm x})$ at hand, we can use the ansatz~\cite{Schiller2000}
\begin{equation}
P(E_\gamma,E_{\rm x}) \propto \rho(E_f) \mathcal{T}(E_\gamma)
\label{eq:ansatz}
\end{equation}
to simultaneously extract the level density $\rho(E_f)$ at the final excitation energy $E_f = E_{\rm x} - E_\gamma$ and the $\gamma$-ray transmission coefficient $\mathcal{T}(E_\gamma)$. 
For dipole decay, the $\gamma$-ray transmission coefficient is connected to the $\gamma$-ray strength function $f(E_\gamma)$ through the expression $\mathcal{T}(E_\gamma) = 2\pi E^3_\gamma f(E_\gamma)$.
The application of Eq.~(\ref{eq:ansatz}) assumes that the generalized form of the BA hypothesis holds. 
Given this expression, both $\rho$ and $\mathcal{T}$ can be extracted from a $\chi^2$ minimization of a chosen area of the primary $\gamma$-ray matrix~\cite{Schiller2000}. 
For $^{120}$Sn, the area confined by the blue lines (area 3) in Fig.~\ref{fig:fgmatrix} was chosen for the decomposition. 
The minimization yields the functional forms of both the $\rho(E_f)$ and $f(E_\gamma)$, except for the absolute value and the slope (see the Supplemental Material for details). 
The level density at low excitation energies is normalized using available information on low-lying discrete levels, while the value $\rho(S_n)$, obtained from the $s$-wave neutron resonance spacing $D_0$ or from systematics, is used to further constrain the normalization. 
Finally, the GSF is normalized to the value of the average total radiative width from $s$-wave neutron resonance experiments. 
Details of the normalization procedure, a presentation of all parameters as well as the choice of the primary $\gamma$-ray matrix area for $^{116,120,124}$Sn can be found in the Supplemental Material. 

\textit{Results and discussion.}$-$Figure~\ref{fig:gammaSF} compares the GSFs for $^{116,120,124}$Sn extracted using the Oslo method (blue) and from inelastic proton scattering \cite{Bassauer2020b} (orange).
In the energy regions where both results overlap, the two fundamentally different methods yield agreement of the $\gamma$-ray energy dependence as well as absolute values within the estimated uncertainty bands for all three nuclei in support of the BA hypothesis. 

Peak-like structures at $E_{\gamma} \approx 6.5$ MeV with a width of about 200 keV (FWHM) are systematically observed in the ($p,p^{\prime}$) data \cite{Bassauer2020b,Krumbholz2015} as highlighted in the lower part of Fig.~\ref{fig:gammaSF}.
The strength at the peak shows an increase from $ 1 \times 10^{-7}$ MeV$^{-3}$ to $2.2 \times 10^{-7}$ MeV$^{-3}$ with the increasing neutron number from $^{116}$Sn to $^{124}$Sn.
While the present as well as $(\gamma,\gamma^\prime)$ data discussed below represent the isovector response, a concentration of isoscalar $E1$ strength has also been found in $^{124}$Sn between 6 and 7 MeV \cite{Endres2012,Pellegri2014}.
The mutual observation in reactions probing the isoscalar and isovector response is considered a signature of the PDR \cite{Savran2013,Bracco2019}.

The $(\gamma,\gamma^\prime)$ data on $^{116,120,124}$Sn \cite{Govaert1998,Oezel-Tashenov2014} show corresponding peaks at about 6.5 MeV with comparable strength but a rather dramatic suppression at higher excitation energies reaching an order of magnitude at 8.5 MeV.
The differences have been attributed to an increasing complexity of the wave functions of the excited states (as expected for the IVGDR), resulting in small branching ratios to the ground state.
Such an interpretation is corroborated by a recent $(\gamma,\gamma^\prime)$ experiment on $^{120}$Sn with improved sensitivity \cite{Muescher2020}.
Applying statistical model corrections of the branching ratios the deduced photoabsorption cross sections agree within the considerable model dependence of such a procedure discussed above.
We note that the recent realization of  $(\gamma,\gamma^\prime\gamma^{\prime\prime})$ experiments in combination with quasi-monoenergetic photon beams from laser Compton backscattering promises a competitive extraction of photoabsorption cross sections below threshold in the future \cite{Isaak2021}.

The possible observation of a peak around 6.5 MeV in the Oslo data is unclear. 
No such structure is visible for $^{116}$Sn.
However, the statistics in the high-$E_{\gamma}$ range for this older experiment were insufficient to track its existence.
A peak at 6.5 MeV can be seen in the GSF of $^{120}$Sn, but the fluctuations of data points below and above are of similar size.
In $^{124}$Sn, where the peak is most pronounced in the ground-state photoabsorption experiments, a potential structure with respect to the uncertainties from the extraction of primary $\gamma$-rays is observed.
We note that although systematic uncertainties are large for $^{124}$Sn due to the absence of level density information from neutron capture reactions, variations within the total error bars may shift the GSF up or down but the peak around 6.5 MeV remains.

An alternative way to test the BA hypothesis with Oslo-type data is to study the GSF as a function of the initial and final excitation energy as outlined in Ref.~\cite{Guttormsen2016}. 
From the primary $\gamma$-ray matrix, we extract the GSF for 256-keV wide excitation-energy bins. 
This way, we can investigate the possible variation of the GSF as a function of  initial excitation energy. 
%
\begin{figure}
\includegraphics[width=1.0\columnwidth]{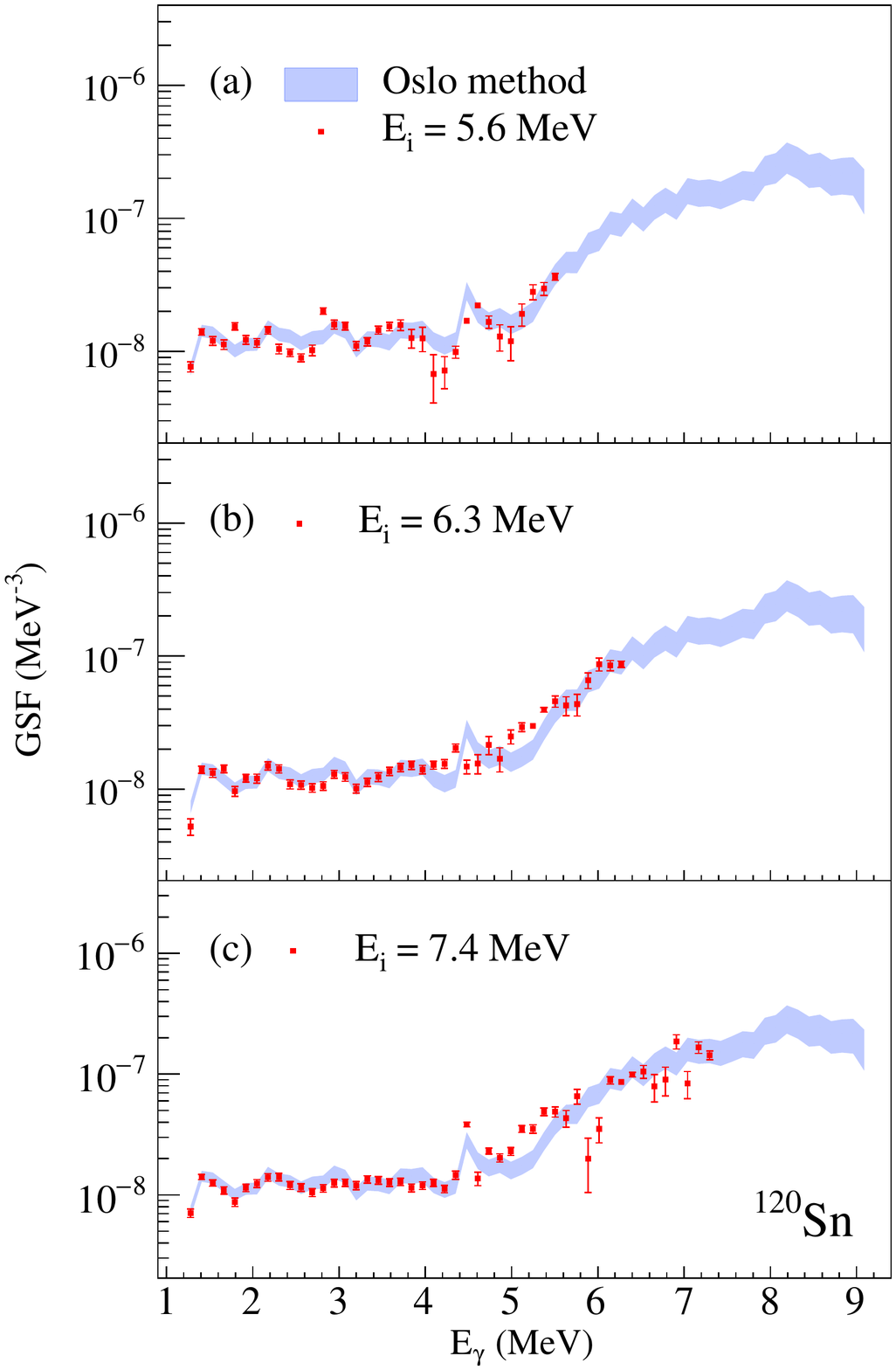}
\caption{\label{fig:wide}
GSF of $^{120}$Sn for several narrow initial excitation energy bins with width of 256 keV (red data points)  compared to the Oslo method result for the full excitation energy range $4.5 \leq E_{\rm x}\leq 9.1$ MeV (blue). 
For the Oslo method the total error band is shown. For both approaches, an $E_\gamma$ bin width of 128 keV is used.   
}
\end{figure}
The results of applying this procedure to the $^{120}$Sn data are illustrated in Fig.~\ref{fig:wide}, where the GSFs for three narrow initial excitation energies are compared to the Oslo-method data extracted from the full $E_{\rm x}$ range.
Each GSF was scaled to the Oslo-method results by a $\chi^2$ fit.
There is overall good agreement, but the GSFs for the selected initial energy bins exhibit stronger  fluctuations compared to the standard Oslo method strength. 
This can be traced back to the reduced number of levels in the initial state bins which lead to an increase of fluctuations of the Porter-Thomas intensity distribution expected for statistical decay \cite{PT}. 
An analog analysis of the final-state energy dependence shows comparable agreement. 


Finally, we test the spin independence of the present results by applying a novel approach to extract the energy dependence of the GSF in a largely model-independent way (the so-called \textit{Shape method}~\cite{Wiedeking2020}). 
Here, we use the capability of the OSCAR array to resolve  the decay to the ground state and to the first excited 2$^+$ state studying again the case of the $^{120}$Sn isotope (regions 1 and 2 in Fig.~\ref{fig:fgmatrix}). 
Compared to the much broader spin range contributing to the full Oslo data set, this defines initial spin windows $J = 1$--$3$ and $J = 1$ for levels directly feeding the $2^+$ state and $0^+$ ground state, respectively.
As the dipole GSF is given by~\cite{Bartholomew1972} 
$$f(E_i,E_f,E_\gamma, J_i^\pi) = \frac{\left< \Gamma_\gamma(E_i,E_f,E_\gamma, J_i^\pi) \right>\rho(E_i,J_i^\pi)}{ {E_\gamma^3}},$$
one can extract information on the shape of the $\gamma$ strength from the intensities $N_D$ (defined in Eq.~(13) of Ref.~\cite{Wiedeking2020}) proportional to the average, partial radiative width $ \left< \Gamma_\gamma(E_i,E_f,E_\gamma, J_i^\pi) \right>$  in the diagonals.

\begin{figure}
\includegraphics[width=1.0\columnwidth]{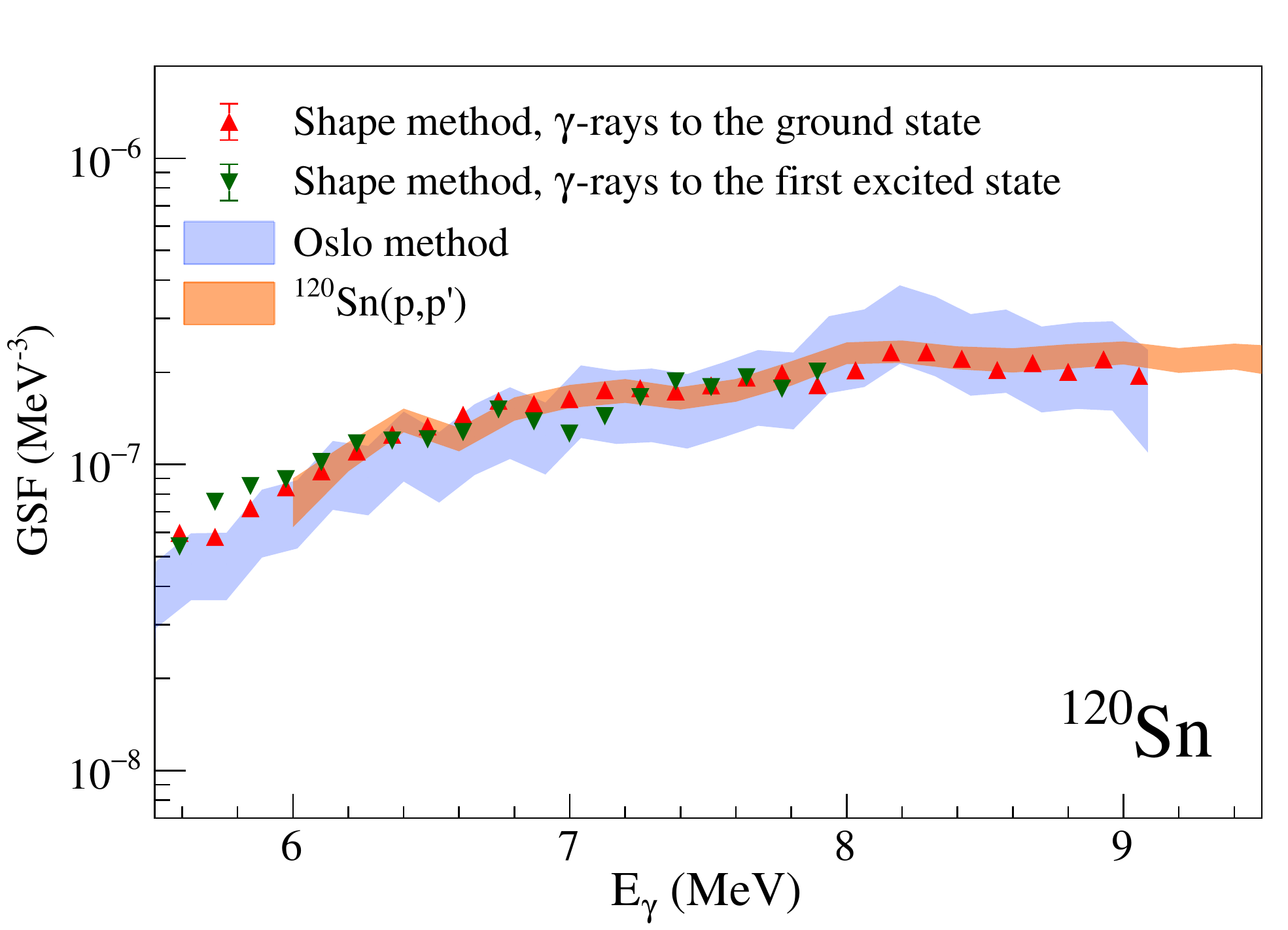}
\caption{\label{fig:comparison} 
Comparison of the GSFs for $^{120}$Sn extracted with the Oslo method (blue band) from selective decay to the ground state and the first excited $2^+$ state utilizing the Shape method (red and green triangles) and from the ($p,p^{\prime}$) data~\cite{Bassauer2020b} (orange band). The  $E_\gamma$ bin widths are 128 keV for the Oslo and Shape-method data and 200 keV for the ($p,p^{\prime}$) data. 
}
\end{figure}
The GSF deduced from the Shape method is shown in Fig.~\ref{fig:comparison} together with those extracted from the Oslo method and from the $(p,p^\prime)$ data.
Data points from decay to the $0^+$ and $2^+$ state are shown by red and green triangles, respectively.
The error bars include only statistical errors, which are typically smaller than the symbol sizes.
Since the Shape method does not provide an absolute normalization of the strength, the results were scaled to the $(p,p^{\prime})$ data by a least-squares fit. 
The shapes of all three GSFs agree within their uncertainties, demonstrating independence from the particular spin distribution of the initial and final states. 
The comparison of the GSF from inelastic proton scattering with the Shape-method data points from ground state decay 
illustrates the direct correspondence between ``upward" and ``downward" strengths. 


\textit{Summary and conclusions.}$-$
We present a critical test of the generalized BA hypothesis in heavy nuclei in the energy region below the neutron threshold. It is based on a comparison of the GSFs in $^{116,120,124}$Sn deduced from relativistic Coulomb excitation in forward-angle inelastic proton scattering \cite{vonNeumannCosel2019} and from Oslo-type experiments.
The two sets of GSFs agree within experimental uncertainties in the energy region between 6 MeV and the neutron threshold demonstrating that the generalized BA
hypothesis holds for the studied cases in this energy region, and experiments based on ground state photoabsorption indeed provide the same information on GSFs in nuclei as Oslo-type experiments.
The presence of peaks around 6.5 MeV attributed to the PDR remains unclear in the Oslo data.  
However, their overall contribution to the GSF -- if present -- is small.
Thus, the assumptions made in the calculations of $(n,\gamma)$ reactions relevant to $r$-process nucleosynthesis are verified.
Further tests of the BA hypothesis include a demonstration of the independence of the GSFs from the energies and spins of initial and final states.
The latter utilizes the novel Shape method \cite{Wiedeking2020} which allows a largely model-independent extraction of the energy dependence of the GSF from the selective decay to specific final states.

It remains an open question to what extent these results can be generalized.
Since we are discussing averaged properties, the most critical parameter is a sufficiently large level density.
The examples studied here are semimagic nuclei with correspondingly low level-density values.
Thus, we expect that our conclusion on the BA hypothesis may hold in general for heavy nuclei with ground state deformation (and thus higher level densities) \cite{Martin2017} except for doubly magic cases \cite{Bassauer2016}.
Future comparisons should explore the limits of ground state  photoabsorption experiments to extract the GSF as a function of $\gamma$ energy, level density, and mass number. 

\begin{acknowledgments}
The authors express their thanks to J.~C.~M\"{u}ller, P.~A.~Sobas, and J.~C.~Wikne at the Oslo Cyclotron Laboratory for operating the cyclotron and providing excellent experimental conditions.
A.~Zilges is thanked for stimulating discussions and providing the $^{120, 124}$Sn targets.
This work was supported in part by the National Science Foundation under Grant No.\ OISE-1927130 (IReNA), by the Deutsche Forschungsgemeinschaft (DFG, German Research Foundation) under Grant No.\ SFB 1245 (project ID 279384907), by the Norwegian Research Council Grant 263030, and by the National Research Foundation of South Africa (Grant No.\ 118846).
A.~C.~L. acknowledges funding by the European Research Council through ERC-STG-2014 under Grant Agreement No.\ 637686, from the “ChETEC” COST Action (CA16117), supported by COST (European Cooperation in Science and Technology), and from JINA-CEE (JINA Center for the Evolution of the Elements) through the National Science Foundation under Grant No.\ PHY-1430152. 

\end{acknowledgments}


\bibliography{tin_2020}

\end{document}